\documentstyle[twocolumn,aps,epsf]{revtex}
\begin{document}
\draft
%
%
%
%
\title{Non-linear response and electron-electron interactions
in mesoscopic metal rings}  
\author{Peter Kopietz\cite{address} and 
Axel V\"{o}lker\cite{address}}
\address{
Institut f\"{u}r Theoretische Physik der Universit\"{a}t G\"{o}ttingen,
Bunsenstrasse 9, D-37073 G\"{o}ttingen, Germany}
\date{\today}
\maketitle
\begin{abstract}

A time-dependent  
electric field
gives rise to a stationary non-equilibrium
current $I^{(2)}$ around a
mesoscopic metal ring threaded by a magnetic flux.
We show that this current,
which is proportional to the intensity of the field,
is closely related to the
exchange part of the interaction contribution
to the {\it{equilibrium}}
persistent current, and that the
corresponding non-linear conductivity
directly measures the weak localization correction
to the polarization.
We explicitly calculate the disorder average 
of $I^{(2)}$ in the diffusive regime 
as function of the frequency  of the electric
field and the static flux  piercing the ring, and
suggest an experiment to test our theory.

\end{abstract}
\narrowtext

\section{Introduction}

Electron-electron interactions in disordered
mesoscopic metals are not very well understood.
The usual perturbative machinery of many-body
theory is not always applicable in these systems, because 
often the intricate interplay between
interactions and disorder in phase coherent systems
cannot be described by means of simple perturbation theory.
The persistent current\cite{Hund38} in a mesoscopic
diffusive metal ring threaded by a magnetic flux  belongs to this
category.
Seven years after the seminal experiment by L\'{e}vy {\it{et al.}}\cite{Levy90},
there seems to be general agreement that 
electron-electron interactions are essential to explain
the surprisingly large
magnitude of the experimentally measured average persistent current
in an array  of $10^7$ Cu-rings.
Note that the experiment by L\'{e}vy {\it{et al.}} has recently been independently
confirmed\cite{Mohanty96}.

In this work we shall study electron-electron interactions 
in mesoscopic metal rings by means of a somewhat
unconventional approach,
which is based on the connection
between electron-electron
interactions on the one hand, and non-linear response 
to an external electromagnetic field on the other hand.
In the context of persistent currents this
connection has recently been pointed out
by Kravtsov and Yudson\cite{Kravtsov93}, who
considered the {\it{time-independent}} part
of the non-equilibrium current proportional to the 
intensity of an external 
longitudinal electric field ${{E}} ( t ) = {\rm Re} [ {{E}} ( \omega )
e^{i \omega t}]$,
  \begin{equation}
 I^{(2)}  = {\rm Re} \left[
 \sigma^{(2)} ( \omega , \phi )  \right] | {{E}}
 ( \omega ) |^2 
 \; .
 \label{eq:Inedef}
 \end{equation}
The so defined non-linear conductivity
$\sigma^{(2)} ( \omega , \phi )$
(see Eqs.(\ref{eq:Ineav}) and (\ref{eq:sigma2av}) below) 
is a function of the frequency
$\omega $ of the external electric field, as well as of
the static flux $\phi$ piercing the ring.
The fact that non-linear response and interactions
are closely related 
becomes obvious in a path-integral approach. 
Indeed, it is well known\cite{Feynman65} that
the Coulomb interaction between electrons
can be obtained by integrating
the exponential of the
coupled Maxwell-matter action 
over the fluctuating quantum electric and magnetic fields.
However, in a path integral approach we may also perform
the integrations in a different order. Thus,
an alternative method to obtain
the equilibrium current 
is to calculate first the non-equilibrium
current 
for a given realization of the electromagnetic fields, 
and then performing an average over these fields.
The effective action for this averaging procedure 
is obtained by integrating first over the electronic
degrees of freedom, keeping the electromagnetic fields fixed.
In this work we shall show that
this procedure leads to new perspectives in the
role of electron-electron interactions for persistent currents,
which can be tested experimentally by means of  non-linear
transport experiments.
We shall also re-examine and correct an earlier  
calculation of the non-equilibrium current (\ref{eq:Inedef})
due to Kravtsov and Yudson\cite{Kravtsov93}.

The equivalence between Coulomb interactions 
and fluctuating electromagnetic fields
has been employed previously by Altshuler, Aronov and
Khmelnitsky\cite{Altshuler82}
in their calculation of the dephasing rate due to electron-electron
interactions in disordered metals.
See also Ref.\cite{Oppen97} for a recent 
study of interaction effects in mesoscopic
conductors with the help of this approach.

The plan of the rest of this paper is as follows: In Sec.\ref{sec:int}
we shall use well-known functional techniques 
to show how the above averaging procedure can be carried out in
practice,
and elucidate the precise connection between
non-linear response and electron-electron interactions.
In particular, we show that the  non-equilibrium current
(\ref{eq:Inedef})
is closely related to the Fock contribution to the equilibrium
persistent current, and that the
non-linear conductivity $\sigma^{(2)} ( \omega , \phi )$ in Eq.(\ref{eq:Inedef})
can be simply obtained from the weak localization correction
to the polarization of the system.
In Sec.\ref{sec:quad} we shall explicitly
evaluate the average non-equilibrium current
$\overline{I}^{(2)}$ as function of
$\omega$ and $\phi$, and compare our result with Ref.\cite{Kravtsov93}.
(Here and below the over-bar denotes averaging over the disorder.)
In Sec.\ref{sec:conc} we discuss 
possibilities to test our theory  experimentally,
and conclude in Sec.\ref{sec:sum} with a brief summary.

\section{From interactions to non-linear response}
\label{sec:int}

In this section we shall map the problem of calculating  the equilibrium
persistent current of interacting electrons onto
an effective non-equilibrium problem in imaginary time.
This is achieved by means of a Hubbard-Stratonovich transformation, a
well known technique in the theory of strongly 
correlated electrons\cite{Kopietz97,Negele88}.

\subsection{Definition of the problem}

We consider a system consisting 
of electrons with charge $-e$ and mass $m$ which are
confined to a thin ring with circumference $L$
and cross section $L_{\bot}^2$, with $L_{\bot} \ll L$.
The electrons interact
with two-body Coulomb forces and move in 
a static random potential $U ( {\bf{r}})$.
Identifying the
position along the circumference of the ring with the
$x$-coordinate, the equilibrium current $I$ around the ring
can be written as
 \begin{equation}
 I = \frac{- e}{L} \int d {\bf{r}} j ( {\bf{r}})
 \label{eq:Ieqdef}
 \; ,
 \end{equation}
where the integral is over the volume ${\cal{V}} = L L_{\bot}^2$ of the ring, and
the particle current density $j ({\bf{r}})$ (in $x$-direction) 
can be expressed in terms of the exact imaginary time Green's function
$G ( {\bf{r}} , {\bf{r}}^{\prime} , \tau - \tau^{\prime})$ as
 \begin{eqnarray}
 j ( {\bf{r}}) & = &  
 \lim_{\tau^{\prime} \rightarrow \tau }
 \lim_{ {\bf{r}}^{\prime} \rightarrow {\bf{r}}}
 \hat{J}_{ x , x^{\prime} }
 G ( {\bf{r}} , {\bf{r}}^{\prime} , \tau - \tau^{\prime} - 0^{+} )
 \nonumber
 \\
 & = & 
 T \sum_{ n = - \infty}^{\infty} e^{i \tilde{\omega}_n 0^{+}}
 \lim_{ {\bf{r}}^{\prime} \rightarrow {\bf{r}}}
 \hat{J}_{ x , {x^{\prime}} }
 G ( {\bf{r}} , {\bf{r}}^{\prime} ,  i \tilde{\omega}_n )
 \; , 
 \label{eq:curdensdef}
 \end{eqnarray}
with the differential
operator
 \begin{equation}
 \hat{J}_{ {x} , {{x^{\prime}}}} \equiv 
 \frac{1}{2m i} \left( \partial_x - \partial_{x^{\prime}}
 \right) +  \frac{a}{m}
 \; \; \; , \; \; a = \frac{2 \pi  }{ L  } \frac{\phi}{\phi_0}
 \label{eq:jdifdef}
 \; .
 \end{equation}
Here $\phi_0 = 2 \pi c /e$ is the flux quantum, and we have
introduced the imaginary frequency Fourier transform of the 
Green's function,
 \begin{equation}
 G ( {\bf{r}} , {\bf{r}}^{\prime} , \tau - \tau^{\prime})
 = T \sum_{n = - \infty}^{\infty} e^{ - i \tilde{\omega}_n (\tau -
 \tau^{\prime})}
 G ( {\bf{r}} , {\bf{r}}^{\prime} ,  i \tilde{\omega}_n )
 \; ,
 \end{equation}
where $\tilde{\omega}_n = 2 \pi ( n + \frac{1}{2}) T$.
For simplicity, we shall work with spinless electrons and use units
where
$\hbar$ and the Boltzmann-constant are set equal to unity.
This amounts to measuring  temperatures $T$
and frequencies $\omega$ in units of energy.
The Green's function at constant chemical
potential $\mu$ can be represented as a 
functional integral over Grassmann fields $\psi$ and $\psi^{\dagger}$
in the usual way\cite{Negele88},
 \begin{equation}
 G ( {\bf{r}} , {\bf{r}}^{\prime} , \tau - \tau^{\prime})
 = - \frac{ \int {\cal{D}} \left\{ \psi \right\} 
 e^{- S \left\{ \psi \right\}}  \psi ( {\bf{r}} , \tau )
 \psi^{\dagger} ( {\bf{r}}^{\prime} , \tau^{\prime} )}{
 \int {\cal{D}} \left\{ \psi \right\} 
 e^{- S \left\{ \psi \right\}} }
 \label{eq:Gfunc}
 \; ,
 \end{equation}
where $S \{ \psi \} = S_0 \{ \psi \} +
S_{\rm int} \{ \psi \}$, with
 \begin{eqnarray}
 S_0 \{ \psi \} & = & - \frac{1}{T} \sum_{k k^{\prime}} \psi^{\dagger}_k 
 [\hat{G}_0^{-1}]_{k k^{\prime}} \psi_{k^{\prime}}
 \label{eq:S0def} 
 \; ,
 \\
 S_{\rm int} \{ \psi \} & = & \frac{1}{2T} \sum_{q} f_{\bf{q}} \rho_{-q} \rho_q
 \label{eq:Sint}
 \; .
 \end{eqnarray}
Here
$\rho_q = \sum_k \psi^{\dagger}_k \psi_{k+q}$, and
the inverse non-interacting Green's function matrix in the
momentum-frequency
basis for a given realization of the disorder
potential is
 \begin{equation}
 [ \hat{G}_0^{-1}]_{k k^{\prime}} = \delta_{ k k^{\prime}} 
  \left[ i \tilde{\omega}_n - \frac{
 ( {\bf{k}} + {\bf{a}})^2 }{2m} + \mu \right] - \delta_{n n^{\prime}} U_{ {\bf{k}} -
 {\bf{k}}^{\prime}} 
 \; ,
 \label{eq:Gkdef}
 \end{equation}
where ${\bf{a}}$  is a vector potential directed along the
circumference of the ring (which in our convention is
identified with the $x$-direction), with magnitude 
$a \equiv | {\bf{a}} |  = \frac{2 \pi}{L} \frac{\phi}{\phi_0}$.
For simplicity 
we have introduced collective labels
$k = [{\bf{k}} , i \tilde{\omega}_n ]$ and $q = [ {\bf{q}} , i
\omega_n]$ 
for wave-vector and Matsubara frequencies, where $\tilde{\omega}_n =
2 \pi (n + \frac{1}{2}) T $ is a fermionic frequency, and $\omega_n =
2 \pi n T$ is a bosonic one.
The Grassmann variables $\psi_k $ are the Fourier components of
the field $\psi ( {\bf{r}} , \tau )$, i.e.
$\psi ( {\bf{r}} , \tau ) = {\cal{V}}^{-1/2} \sum_k e^{i ({\bf{k}} \cdot
  {\bf{r}} - \tilde{\omega}_n \tau )} \psi_k$.
The Fourier transforms $U_{\bf{q}}$ of the disorder potential
and $f_{\bf{q}}$ of the Coulomb potential are normalized such that
both have units of energy\cite{footnote2}, i.e.
 \begin{eqnarray}
 U_{\bf{q}} & = & \frac{1}{{\cal{V}}} \int d{\bf{r}} e^{ - i {\bf{q}} \cdot {\bf{r}}} U
 ( {\bf{r}})
 \label{eq:Uqdef}
 \; , 
 \\
 f_{\bf{q}} & = & \frac{1}{{\cal{V}}^2} \int d {\bf{r}}  d {\bf{r}}^{\prime}
 e^{ - i {\bf{q}} \cdot ( {\bf{r}} - {\bf{r}}^{\prime})} 
 \frac{e^2}{ | {\bf{r}} - {\bf{r}}^{\prime} | }
 \label{eq:fqdef}
 \; .
 \end{eqnarray}
The disorder potential $U ( {\bf{r}})$ is assumed to have
zero average and Gaussian white noise correlations, so that
 \begin{equation}
 \overline{ U_{\bf{q}} U_{{\bf{q}}^{\prime}} } = \tilde{\gamma} \delta_{
   {\bf{q}} , - {\bf{q}}^{\prime} }
 \; ,
 \end{equation}
where the parameter $\tilde{\gamma}$ is a measure for the
strength of the disorder. Within lowest order Born approximation
we may identify
 $\tilde{\gamma}  = \Delta / ( 2 \pi \tau )$,
where $\Delta$ is the average level spacing at the Fermi energy, and
$\tau$
is the elastic lifetime.

The evaluation of the above expression for the current 
would require the solution of the many-body problem in the
presence of disorder, an impossible task.
Perturbative expansions can be performed in powers
of the disorder potential $U_{\bf{q}}$ and in powers
of the Coulomb interaction $f_{\bf{q}}$.
This double expansion is rather subtle.
To obtain sensible results which correctly
take into account 
the physics of diffusion and screening,
infinitely many  powers of $U_{\bf{q}}$ and
$f_{\bf{q}}$ have to be summed. 
In order to make this expansion more transparent and to see the
connection with non-linear response, we shall now map
this problem onto an equivalent problem where the two-body interaction
is replaced by a time-dependent auxiliary field.

\subsection{Hubbard-Stratonovich transformation and equivalent
non-equilibrium problem}
The two-body part $S_{\rm int} \{ \psi \}$ of our effective
action can be decoupled by means of the following Hubbard-Stratonovich
transformation\cite{Kopietz97}
 \begin{eqnarray}
 e^{- S_{\rm int} \{ \psi \}} & = & 
 \nonumber 
 \\
 & & \hspace{-20mm} 
 \frac{ \int {\cal{D}} \{ \Phi \} \exp \left[  - \frac{T}{2} \sum_q
 f^{-1}_{\bf{q}} \Phi_{-q} \Phi_q - i \sum_q \Phi_{-q} \rho_q \right] }{
 \int {\cal{D}} \{ \Phi \} \exp \left[  - \frac{T}{2} \sum_q
 f^{-1}_{\bf{q}} \Phi_{-q} \Phi_q \right] }
 \label{eq:HS}
 \; .
 \end{eqnarray}
Applying this transformation to the denominator and numerator
of Eq.(\ref{eq:Gfunc}), and integrating over the
Grassmann field, the exact current density of the many-body
system can be written as 
 \begin{eqnarray}
 j ( {\bf{r}})
 & = & \frac{ \int {\cal{D}} \{ \Phi\} e^{- S_{\rm eff} \{ \Phi \}} 
 j ( {\bf{r}} , \tau , \{ \Phi \}  ) }{
 \int {\cal{D}} \{ \Phi\} e^{- S_{\rm eff} \{ \Phi \}} }
 \nonumber
 \\
 &  \equiv & \langle j ( {\bf{r}} , \tau , \{ \Phi \}  ) \rangle_{S_{\rm eff}}
 \label{eq:jeq}
 \; .
 \end{eqnarray}
The effective action $S_{\rm eff} \{ \Phi \}$ is given by
 \begin{equation}
 S_{\rm eff} \{ \Phi \} = \frac{T}{2} \sum_{q} f^{-1}_{\bf{q}} \Phi_{-q} \Phi_q
 -   {\rm Tr} \ln \left[ 1 - \hat{G}_0 \hat{V} \right]
 \;  ,
 \label{eq:Seffdef}
 \end{equation} 
where the matrix elements of the infinite matrix $\hat{V}$
are given by $[ \hat{V}]_{k k^{\prime}} = V_{k - k^{\prime}} =
i T \Phi_{k - k^{\prime}}$. 
The quantity
$j ( {\bf{r}} , \tau , \{ \Phi \}  )$
is the non-equilibrium current density for a frozen configuration
of the Hubbard-Stratonovich field, i.e.
 \begin{equation}
 j ( {\bf{r}} , \tau , \{ \Phi \}  ) =
 \lim_{ {\bf{r}}^{\prime} \rightarrow {\bf{r}}}
 \hat{J}_{ x , x^{\prime} }
 {\cal{G}} ( {\bf{r}} , {\bf{r}}^{\prime} , \tau ,  \tau + 0^{+} ) 
 \; .
 \label{eq:jnedef}
 \end{equation}
Here ${\cal{G}}$ satisfies the partial differential equation
 \begin{eqnarray}
 \left[ - \partial_{\tau} - \frac{ ( - i \nabla_{\bf{r}} + {\bf{a}} )^2}{2m} + \mu - U (
 {\bf{r}}) - V ( {\bf{r}} , \tau ) \right]
 {\cal{G}} ( {\bf{r}} , {\bf{r}}^{\prime} , \tau ,  \tau^{\prime}   )  &  & 
 \nonumber
 \\
 & & \hspace{-60mm}
 = \delta ({\bf{r}} - {\bf{r}}^{\prime})
 \delta^{\ast} ( \tau - \tau^{\prime})
 \; ,
 \label{eq:Gnedef}
 \end{eqnarray}
where the time-dependent potential $V ( {\bf{r}} , \tau )$
is defined by
 \begin{equation}
 V ( {\bf{r}} , \tau ) 
 = \sum_{q} e^{i ({\bf{q}} \cdot {\bf{r}} - \omega_n \tau )}
 V_q \; \; \; , \; \; \; V_q = i T \Phi_q \; ,
 \label{eq:Vrt}
 \end{equation}
and $\delta^{\ast} ( \tau ) = T \sum_n e^{- i \tilde{\omega}_n \tau }$
is the antiperiodic imaginary time $\delta$-function.
Note that the potential $V ( {\bf{r}} , \tau )$ is a periodic function
of $\tau$, i.e.
$V ( {\bf{r}} , \tau + 1/T) = V ( {\bf{r}} , \tau )$.
On the other hand, the fermionic Green' s function ${\cal{G}}$ has to satisfy
antiperiodic boundary conditions in each imaginary time 
variable\cite{Kadanoff62},
 \begin{eqnarray}
 {\cal{G}} ( {\bf{r}} , {\bf{r}}^{\prime} , \tau + 1/T ,  \tau^{\prime}  ) & = & - 
 {\cal{G}} ( {\bf{r}} , {\bf{r}}^{\prime} , \tau ,  \tau^{\prime}  ) 
 \nonumber 
 \\
 & =  & 
 {\cal{G}} ( {\bf{r}} , {\bf{r}}^{\prime} , \tau ,  \tau^{\prime} + 1/T)
 \; .
 \label{eq:antibc}
 \end{eqnarray}
The above transformation is exact, and allows us to
clarify the precise connection between interactions
and non-linear response\cite{Kravtsov93}. 
Note that Eq.(\ref{eq:Gnedef}) defines the
{\it{imaginary time}} 
non-equilibrium Green's function of  non-interacting
fermions subject to an external imaginary time potential 
$V ( {\bf{r}} , \tau )$. Of course, for a comparison with
experiments, which measure the
non-linear response to external fields, 
we need to know the real time dynamics.
While within linear response the well-known fluctuation-dissipation theorem
tells us how to obtain the real time response by simple analytic
continuation from an imaginary time formalism, in the case of
non-linear
response the situation is more complicated. 
Nevertheless, even then the analytic continuation
from the imaginary time response to real times is possible, 
provided the time-dependence of the external potential
can be analytically continued, and
the potential is adiabatically switched on\cite{Voelker97}.
This point,
which apparently is not widely appreciated in the literature, 
has already been discussed in the classic textbook by
Kadanoff and Baym\cite{Kadanoff62}.

\subsection{How functional averaging reproduces the equilibrium current}

\begin{figure}
\epsfysize3cm 
\hspace{5mm}
\epsfbox{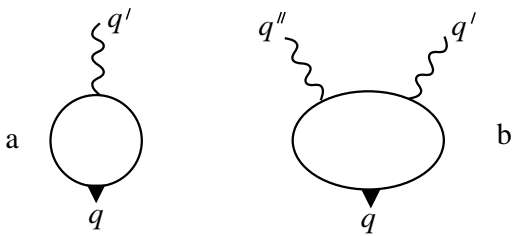}
\vspace{5mm}
\caption{
(a) Graphical representation of the current-density $j^{(1)}_q$ in
linear response, see Eq.(\ref{eq:j1}).
The solid lines represent non-interacting Green's functions
for fixed  disorder potential, the wavy lines
represent the fields $V_q$, and the black triangle denotes the current
vertex $(k_x + a + q_x /2)/m$.
(b) Graphical representation of $j^{(2)}_q$.
}
\label{fig:resp}
\end{figure}

\begin{figure}
\epsfysize4cm 
\hspace{5mm}
\epsfbox{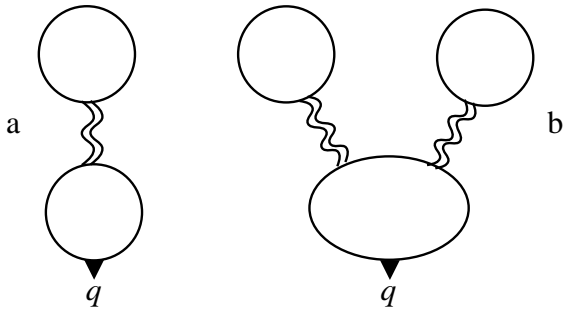}
\vspace{5mm}
\caption{Hartree contributions to the 
functional average of the non-equilibrium
current density given in Eqs.(\ref{eq:jFT}) and (\ref{eq:jqexp}).
(a) Contribution from 
the linear response current $j^{(1)}_q$ (b) 
Contribution from the quadratic response $j^{(2)}_q$.
The double wavy line is the RPA interaction.
}
\label{fig:hartree}
\end{figure}

\begin{figure}
\epsfysize3cm 
\hspace{5mm}
\epsfbox{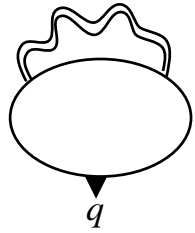}
\vspace{5mm}
\caption{
Fock contribution to functional average of
$j^{(2)}$.
}
\label{fig:fock}
\end{figure}

For a calculation of
the equilibrium persistent current to first order in the
RPA (random phase approximation) screened interaction, it is sufficient
to expand the effective action (\ref{eq:Seffdef})  and the
non-equilibrium current-density $j ( {\bf{r}} , \tau , \{ \Phi \})$
defined in Eq.(\ref{eq:jnedef}) to second order
in the Hubbard-Stratonovich field. The
resulting Gaussian integrations can then be performed exactly.
In this approximation the effective action is given by
 \begin{eqnarray}
 S_{\rm eff} \{ \Phi \} & \approx &
 i \sum_{{q}} N_0 ( q ) \Phi_{- {q}} 
 \nonumber 
 \\
 &  & \hspace{-15mm} +
 \frac{T}{2 } \sum_{q q^{\prime}} \left[ \delta_{qq^{\prime}} f^{-1}_{\bf{q}}  +
 \Pi_0 ( q , q^{\prime} ) \right]
 \Phi_{-q } \Phi_{q^{\prime}} + \ldots
 \label{eq:Seffexpansion}
 \; \; ,
 \end{eqnarray}
where
 \begin{eqnarray} 
 N_0 ( {{q}} )  & = & \delta_{n0} N_0 ({\bf{q}})
 = T \sum_k [ \hat{G}_0 ]_{k+q , k} 
 \label{eq:N0def}
 \; ,
 \\
 \Pi_0 ( q , q^{\prime} ) 
 & = & \delta_{ n n^{\prime}} \Pi_0 ( {\bf{q}} , {\bf{q}}^{\prime} , i
 \omega_n)
 \nonumber
 \\
 & = &
 -  T \sum_{k k^{\prime}}
 [\hat{G}_0]_{k+q,k^{\prime} + q^{\prime}} [\hat{G}_0]_{k^{\prime}k}
 \label{eq:pol}
 \;  .
 \end{eqnarray}
Physically
$ N_0 ( {\bf{q}} )$ 
is the spatial Fourier component of the density, and
$ \Pi_0 ( {\bf{q}} , {\bf{q}}^{\prime} , i \omega_n )$
is the non-interacting polarization\cite{footnote2} for a given realization
of the disorder potential $U ( {\bf{r}})$.
To expand the non-equilibrium current-density, it is
convenient to consider the Fourier components, 
 \begin{equation}
 j ( {\bf{r}} , \tau , \{ \Phi \}) = \frac{1}{{\cal{V}}} \sum_q e^{ i (
 {\bf{q}} \cdot {\bf{r}} - \omega_n \tau)} j_q 
 \label{eq:jFT}
 \; .
 \end{equation}
Note that with this normalization the equilibrium current
defined in Eq.(\ref{eq:Ieqdef}) is simply given by
 \begin{equation}
 I = \frac{-e}{L} \langle j_{q=0} \rangle_{S_{\rm eff}}
 \label{eq:Ieqwrite}
 \;  ,
 \end{equation}
where $q = 0$ means ${\bf{q}} = 0$ and $\omega_n = 0$.
For $j_q $ we obtain the following expansion in powers of $V_q = i T \Phi_q$,
 \begin{equation}
 j_q = j_q^{(0)} + j_q^{(1)} + j_q^{(2)} + \ldots
 \; ,
 \label{eq:jqexp}
 \end{equation}
where
 \begin{eqnarray}
 j_q^{(0)} & = & T \sum_k \frac{ {k}_x + a + {q_x}/{2} }{m} [\hat{G}_0]_{k+q,k}
 \label{eq:j0}
 \; ,
 \\
 j_q^{(1)} & = & \sum_{q^{\prime}} K^{(1)} ( q , q^{\prime})
 V_{q^{\prime}}
 \label{eq:j1}
 \; ,
 \\
 j_q^{(2)} & = & \sum_{q^{\prime} q^{\prime \prime}} K^{(2)} ( q ,
 q^{\prime} , q^{\prime \prime})
 V_{q^{\prime}} V_{ q^{\prime \prime}}
 \label{eq:j2}
 \; .
 \end{eqnarray}
The linear response function $K^{(1)} ( q , q^{\prime})$ 
can be identified with the non-interacting
correlation function between density and current-density,
 \begin{equation}
 K^{(1)} ( q , q^{\prime}) = T \sum_{k k^{\prime}}
 \frac{{k}_x + a + {q_x}/{2} }{m}
 [ \hat{G}_0]_{k +q, k^{\prime} + q^{\prime}}
 [ \hat{G}_0]_{k^{\prime}, k}
 \; .
 \label{eq:K1}
 \end{equation}
The quadratic response function is
 \begin{eqnarray}
 K^{(2)} ( q , q^{\prime} , q^{\prime \prime}) & = & T \sum_{k
 k^{\prime} k^{\prime \prime}}
 \frac{{k}_x + a + {q_x}/{2}}{m} 
 [ \hat{G}_0]_{k +q, k^{\prime} + q^{\prime}}
 \nonumber
 \\
 & \times &
 [ \hat{G}_0]_{k^{\prime}, k^{\prime \prime}}
 [ \hat{G}_0 ]_{ k^{\prime \prime} - q^{\prime \prime}, k}
 \; .
 \label{eq:K2}
 \end{eqnarray}
Graphical representations of $j^{(1)}_q$ and $j^{(2)}_q$ are shown in
Figs.\ref{fig:resp}(a) and (b).
%
%
%
%
%
%
%
%
It is instructive to see how functional averaging of these expressions
with the effective action (\ref{eq:Seffexpansion})
yields the well-known\cite{Ambegaokar90,Kopietz97b} 
interaction corrections to the equilibrium
current to first order in the RPA interaction. 
Of course, the equilibrium current is more easily obtained from the
derivative of the thermodynamic potential with respect
to the static flux\cite{Ambegaokar90,Kopietz97b}, 
but the following calculation clarifies the
close connection between non-linear response and electron-electron
interactions\cite{Kravtsov93}.
To perform the  Gaussian integration, it is convenient to first
eliminate the linear term in Eq.(\ref{eq:Seffexpansion}) by redefining
the $\Phi$-field such that its Gaussian average vanishes. 
This is achieved with the help of the shift-transformation
$\Phi_q =  \tilde{\Phi}_q - i T^{-1} \sum_{q^{\prime}} f^{\rm RPA}_{ q  q^{\prime} } 
N_0 ( q^{\prime } )$, or
equivalently for $V_q = i T \Phi_q$,
 \begin{equation}
 V_q = \tilde{V}_q + 
 \sum_{q^{\prime}} f^{\rm RPA}_{ q  q^{\prime} } 
 N_0 ( q^{\prime } )
 \label{eq:Vshift}
 \; .
 \end{equation}
Here $f^{\rm RPA}_{q q^{\prime}}$ is the
{\it{inverse}} of the infinite matrix with elements
$\delta_{q q^{\prime}} f^{-1}_{\bf{q}} + \Pi_0 ( q , q^{\prime} )$.
It follows that within the Gaussian approximation
 \begin{eqnarray}
 \langle \tilde{V}_q \rangle_{S_{\rm eff}} & = & 0
 \label{eq:V1av}
 \; ,
 \\
 \langle \tilde{V}_{q} \tilde{V}_{-q^\prime} \rangle_{S_{\rm eff}} & = & 
 - T f^{\rm RPA}_{ q q^{\prime} }
 \label{eq:V2av}
 \; .
 \end{eqnarray}
Substituting Eq.(\ref{eq:Vshift}) into Eqs.(\ref{eq:j1}) and
(\ref{eq:j2}),
and averaging over the $\tilde{V}$-field in Gaussian approximation,
it is now easy to show  
 \begin{eqnarray}
 \langle j_q^{(1)} \rangle_{S_{\rm eff}} & = & \sum_{q^{\prime} }
 \sum_{q_1} K^{(1)} ( q , q^{\prime})
 f^{\rm RPA}_{q^{\prime} q_1} N_0 ( q_1)
 \label{eq:j1av}
 \; ,
 \\
 \langle j_q^{(2)} \rangle_{S_{\rm eff}} & = & \sum_{q^{\prime}
 q^{\prime \prime}} \sum_{q_1 q_2} 
 K^{(2)} ( q ,
 q^{\prime} , q^{\prime \prime})
 f^{\rm RPA}_{q^{\prime} q_1 } f^{\rm RPA}_{q^{\prime \prime} q_2 } 
 N_0 ( q_1 ) N_0 ( q_2)
 \nonumber
 \\
 & - &
 T \sum_{q^{\prime}
 q^{\prime \prime}}  
 K^{(2)} ( q ,
 q^{\prime} , - q^{\prime \prime}) f^{\rm RPA}_{q^{\prime} q^{\prime \prime}} 
 \label{eq:j2av}
 \; .
 \end{eqnarray}
Graphically Eq.(\ref{eq:j1av}) and the first term
in Eq.(\ref{eq:j2av}) can be represented by the Hartree
diagrams shown in Fig.\ref{fig:hartree}, while the
second term in Eq.(\ref{eq:j2av}) is represented by the Fock 
diagram in Fig.\ref{fig:fock}.
%
%
%
%
%
%
%
%
To see that for $q=0$ Eqs.(\ref{eq:j1av}) and (\ref{eq:j2av}) reduce to
the well-known\cite{Ambegaokar90,Kopietz97b} 
first order (in the RPA interaction) corrections  to the
equilibrium current, 
we use the following exact identity
 \begin{eqnarray}
 \sum_{{\bf{k}}^{\prime}} G_0 ( {\bf{k}} , {\bf{k}}^{\prime} , i
 \tilde{\omega}_n) \frac{ k_x^{\prime} + a }{m} G_0 ( {\bf{k}}^{\prime} ,
 {\bf{k}}^{\prime \prime} ,
 i \tilde{\omega}_n ) & &
 \nonumber
 \\
 & & \hspace{-40mm } =
 \frac{L \phi_0}{2 \pi} \frac{\partial}{\partial
 \phi} G_0 (  {\bf{k}} , {\bf{k}}^{\prime \prime} , i \tilde{\omega}_n )
 \; ,
 \label{eq:identity1}
 \end{eqnarray}
where $[\hat{G}_0]_{k k^{\prime}} =
\delta_{n n^{\prime}} G_0 ( {\bf{k}} , {\bf{k}}^{\prime} ,
i \tilde{\omega}_n )$.
Eq.(\ref{eq:identity1}) can be easily proven by taking the
derivative of both sides of
Eq.(\ref{eq:Gnedef}) (with $V ( {\bf{r}} , \tau )$ set equal to zero) with respect to $a$.
With the help of this identity
we obtain for the functional average of the linear response
current (\ref{eq:j1av})
 \begin{equation}
 \frac{-e}{L} \langle j^{(1)}_0 \rangle_{S_{\rm eff}} = - \frac{c}{2}
 \sum_{ {\bf{q}} {\bf{q}}^{\prime}} f^{\rm RPA}_{ {\bf{q}}
 {\bf{q}}^{\prime} } \frac{\partial}{\partial \phi } \left[
 N_0 ( - {\bf{q}}) N_0 ({\bf{q}}^{\prime}) \right]
 \; ,
 \label{eq:j1avav}
 \end{equation}
where $f^{\rm RPA}_{ {\bf{q}} {\bf{q}}^{\prime}} \equiv
f^{\rm RPA}_{ {\bf{q}} 0, {\bf{q}}^{\prime} 0}$ is the static RPA
interaction.
Similarly, for $q=0$ Eq.(\ref{eq:j2av}) can be written in the form
 \begin{equation}
 \frac{-e}{L} \langle j^{(2)}_0 \rangle_{S_{\rm eff}}  =  - \frac{c}{2}
 \sum_{ {\bf{q}} {\bf{q}}^{\prime}} \left[ 
 \frac{ \partial}{\partial \phi} f^{\rm RPA}_{ {\bf{q}}
 {\bf{q}}^{\prime} } \right] 
 N_0 ( - {\bf{q}}) N_0 ( {\bf{q}}^{\prime})  + I_F
 \; ,
 \label{eq:j2avav}
 \end{equation}
where
 \begin{equation}
 I_F =  - \frac{c}{2} \sum_{ {\bf{q}} {\bf{q}}^{\prime}} T \sum_n
 f^{\rm RPA}_{ {\bf{q}}  i {\omega}_n , {\bf{q}}^{\prime}
 i {\omega}_n } \frac{ \partial}{\partial \phi}
 \Pi_0 ( {\bf{q}} , {\bf{q}}^{\prime} , i {\omega}_n)
 \label{eq:IF}
 \end{equation}
is the
Fock contribution to the equilibrium persistent
current\cite{Ambegaokar90,Kopietz97b}.
The sum of Eq.(\ref{eq:j1avav}) and the first term in
Eq.(\ref{eq:j2avav})
yield the (non self-consistent) Hartree contribution to the
equilibrium current,
 \begin{equation}
 I_H = - \frac{c}{2} \frac{\partial}{\partial \phi }
 \sum_{ {\bf{q}} {\bf{q}}^{\prime}} f^{\rm RPA}_{ {\bf{q}}
 {\bf{q}}^{\prime} }
 N_0 ( - {\bf{q}}) N_0 ( - {\bf{q}}^{\prime})
 \; .
 \label{eq:IH}
 \end{equation}
We have argued elsewhere\cite{Kopietz93} that the neglect of self-consistency
in Eq.(\ref{eq:IH}) does not properly take into account the
subtle interplay between disorder and interactions,
so that the correct order of magnitude of the Hartree current
can only be obtained by means 
of a self-consistent calculation.

\section{Quadratic response to an external electric field}
\label{sec:quad}

\subsection{Derivation of the non-equilibrium current from the
Fock correction to the equilibrium current}

Our rather unconventional derivation of the interaction
correction to the equilibrium current
makes the connection between electron-electron interactions and
non-linear response manifest. In fact, from our derivation
it is clear that the Fock contribution (\ref{eq:IF}) to the 
equilibrium current is closely related
to the non-equilibrium current given in Eq.(\ref{eq:Inedef}).
Physically our Hubbard-Stratonovich field
$\Phi$ can be identified with the scalar potential
of electromagnetism, which is generated self-consistently 
by the motion of the electrons\cite{Feynman65,Kopietz97}.
Therefore the negative gradient of our auxiliary potential
$V ( {\bf{r}} , \tau ) $ is the
effective force acting on the electrons, which in turn
can be associated with an internal electric field
${\bf{E}} ( {\bf{r}} , \tau )$,
 \begin{equation}
 -e {\bf{E}} ( {\bf{r}} , \tau )  = - \nabla V ( {\bf{r}} , \tau )
 \; .
 \label{eq:forcedef}
 \end{equation}  
Defining ${\bf{E}} ( {\bf{r}} , \tau) = \sum_q e^{ i ( {\bf{q}} \cdot
  {\bf{r}} -
\omega_n \tau )} {\bf{E}}_q$ and using Eq.(\ref{eq:Vrt}), we have
$e {\bf{E}}_q = i {\bf{q}} V_{{q}}$, or
 \begin{equation}
 V_{q} = - i e \frac{ \hat{\bf{ q}} \cdot {\bf{E}}_q }{ | {\bf{q}}|}
 \label{eq:VqE}
 \; ,
 \end{equation}
where ${\hat{\bf{q}}} = {\bf{q}}/ | {\bf{q}}|$.
From Eq.(\ref{eq:V2av}) we thus conclude
 \begin{equation}  
 T f^{\rm RPA}_{q  q^{\prime}} =
 - \frac{e^2}{ | {\bf{q}} | | {\bf{q}}^{\prime} |}
 \langle ( \hat{\bf{q}} \cdot {\bf{E}}_q )  
  ( \hat{{\bf{q}}}^{\prime} \cdot {\bf{E}}_{-q^{\prime}} ) \rangle_{
 S_{\rm eff}} 
 \; ,
 \label{eq:EqEq}
 \end{equation}
so that the Fock contribution (\ref{eq:IF}) to the equilibrium current can
be written as
 \begin{equation}
 I_F = \sum_{q q^{\prime} }
 \sigma^{(2)} ( {\bf{q}} , {\bf{q}}^{\prime} , i \omega_n )
 \langle  ( \hat{\bf{q}} \cdot {\bf{E}}_{q})  
  ( \hat{{\bf{q}}}^{\prime} \cdot {\bf{E}}_{-q^{\prime}} ) \rangle_{S_{\rm eff}}
 \; ,
 \label{eq:IFrewrite}
 \end{equation}
where the non-linear conductivity 
$\sigma^{(2)} ( {\bf{q}} , {\bf{q}}^{\prime} , i \omega_n )$ is given by
 \begin{equation}
  \sigma^{(2)} ( {\bf{q}} , {\bf{q}}^{\prime} , i \omega_n )  = 
\frac{c}{2} 
 \frac{e^2}{ | {\bf{q}} | | {\bf{q}}^{\prime} |}
 \frac{ \partial}{\partial \phi}
 \Pi_0 ( {\bf{q}} , {\bf{q}}^{\prime} , i {\omega}_n)
 \; .
 \label{eq:sigmanldef}
 \end{equation}
Note that functional averaging restores translational invariance in
time,
so that the average in Eq.(\ref{eq:IFrewrite}) is proportional to
$\delta_{n n^{\prime}}$.
This equation shows that
the Fock current can be viewed as the sum of functionally
averaged non-equilibrium currents, generated in second 
order in the internal electric fields associated
with the motion of the electrons. Clearly, the corresponding
non-linear response function $\sigma^{(2)}$ is a system property
that should be independent of the origin of the electric fields.
In particular, if we add an external electric field,
Eq.(\ref{eq:IFrewrite}) is still valid provided
we identify ${\bf{E}}$ with the total electric field. 
Thus, after performing in Eq.(\ref{eq:sigmanldef}) the usual analytic
continuation, 
$i \omega_n \rightarrow
\omega + i 0^{+}$, we conclude that 
the non-linear conductivity $\sigma^{(2)} ( \omega , \phi )$
defined in Eq.(\ref{eq:Inedef}) can be identified with
 \begin{equation}
 \sigma^{(2)} ( \omega , \phi ) = \lim_{ {\bf{q}} , {\bf{q}}^{\prime}
   \rightarrow 0} \sigma^{(2)} ( {\bf{q}} , {\bf{q}}^{\prime} , \omega
   + i 0^{+} )
 \; .
 \label{eq:sigma2iden}
 \end{equation}
Recall that $\sigma^{(2)} ( \omega , \phi )$ 
describes the {\it{time-independent}} part of the non-equilibrium
current that is proportional to the intensity of a 
{\it{time-dependent}}, spatially uniform electric field.

Two remarks are in order. The first concerns the analytic continuation
of the imaginary frequency response to real frequencies.
Because Eq.(\ref{eq:sigmanldef}) relates a
non-linear
response function to the flux-derivative of a
linear response function (the polarization),
we may 
rely on the fluctuation-dissipation theorem to relate real and
imaginary time response. 
Of course, Eq.(\ref{eq:sigmanldef}) can also be obtained
with the help of the identity (\ref{eq:identity1})
from the expression (\ref{eq:K2}) for the general non-linear
response function $K^{(2)} ( q  , q^{\prime} , q^{\prime \prime})$.
It is not difficult to see that quite generally
the correct real frequency response
can be obtained by replacing
$ i \omega_n \rightarrow \omega + i 0^{+}$,
$ i \omega_{n^{\prime}} \rightarrow \omega^{\prime} + i 0^{+}$,
and $ i \omega_{n^{\prime \prime}} \rightarrow 
\omega^{\prime \prime} + i 0^{+}$ for all frequencies\cite{Voelker97}.

Secondly, we would like to emphasize that
so far we have not averaged over the
disorder potential, i.e. all equations given above are valid for an arbitrary
realization of $U ( {\bf{r}})$. 
In the following subsection we shall start from
Eq.(\ref{eq:sigma2iden})  to perform the disorder average
of the non-linear conductivity.
This has the advantage that we only need to average a product of 
{\it{two}}
Green's functions. Note that in the work \cite{Kravtsov93}
the average of the non-linear conductivity has been calculated
directly from  
Eqs.(\ref{eq:j2}) and (\ref{eq:K2}). Because the latter expression
involves a product of {\it{three}} Green's function, it is in this
case more difficult
to identify the dominant disorder diagrams.
One difficulty, which apparently has not been noticed
in Ref.\cite{Kravtsov93}, lies in the fact that 
at long wavelengths and small frequencies density-vertices
are renormalized by singular vertex corrections
involving so-called diffuson propagators\cite{Kopietz97b,Monod92}
This might partially explain the discrepancies between the result of 
Kravtsov and Yudson\cite{Kravtsov93}
and our result discussed below.

\subsection{The relation between 
non-linear conductivity, polarization, and linear conductivity}
\label{subsec:Therel}

Using the fact that 
$\overline{\Pi}_0 ( {\bf{q}} , {\bf{q}}^{\prime} , i {\omega}_n)
= \delta_{ {\bf{q}} {\bf{q}}^{\prime} }
 \overline{\Pi}_0 ( {\bf{q}} , i \omega_n)$, 
we see from Eqs.(\ref{eq:IFrewrite}) and (\ref{eq:sigmanldef})
that the 
disorder averaged static non-equilibrium current that is generated
at quadratic order in a space- and time-dependent longitudinal electric field with
Fourier components $ {\bf{E}} ({\bf{q}} , \omega )$ is given by
 \begin{equation}
 \overline{I}^{(2)}  =
 \overline{\sigma}^{(2)} ( {\bf{q}} , \omega , \phi )
 | {\bf{E}} ( {\bf{q}} ,  \omega ) |^{2}
 \; ,
 \label{eq:Ineav}
 \end{equation}
with
 \begin{equation}
 \overline{\sigma}^{(2)} ( {\bf{q}} , \omega , \phi ) =
 \frac{c}{2} \frac{e^2}{ {\bf{q}}^2 } \frac{ \partial}{\partial \phi}
 \overline{\Pi}_0 ( {\bf{q}} , \omega )
 \; .
 \label{eq:sigma2av}
 \end{equation}
We thus need to know the flux-dependent part of the disorder averaged
polarization of the ring. For frequencies 
$| \omega |
\raisebox{-0.5ex}{$\; \stackrel{<}{\sim} \;$} \Delta$
and wave-vectors 
$ | {\bf{q}} |
\raisebox{-0.5ex}{$\; \stackrel{<}{\sim} \;$} 2 \pi / \ell$
(where $\ell = v_F \tau$ is the elastic mean free path)
non-perturbative methods are necessary to calculate this quantity\cite{Efetov96}.
Here we are interested in the high frequency regime
$| \omega |
\raisebox{-0.5ex}{$\; \stackrel{>}{\sim} \;$} \Delta$, where
we may use the impurity diagram technique\cite{Altshuler85}.
However, for $| \omega | \tau
\raisebox{-0.5ex}{$\; \stackrel{<}{\sim} \;$} 1$
and $| {\bf{q}}|  \ell
\raisebox{-0.5ex}{$\; \stackrel{<}{\sim} \;$} 1$
the direct diagrammatic calculation of
$\frac{\partial}{\partial \varphi} 
\overline{\Pi}_0 ( {\bf{q}} , \omega )$ 
is not so easy, because
there exists non-trivial cancellations between 
vertex corrections to the density vertices\cite{Monod92,Kopietz97b}.
Physically, these corrections arise from the
diffusive motion of the electrons in the disordered metal.
To take these corrections into account  without having to
perform complicated manipulations, we use the
exact relation between irreducible polarization and
longitudinal conductivity $\overline{\sigma} ( {\bf{q}} , \omega )$\cite{Pines89}, which in our
normalization\cite{footnote2}
reads
  \begin{equation}
 \overline{\Pi} ( {\bf{q}} , \omega ) =  i \frac{ {\bf{q}}^2}{
    \omega } \frac{ {\cal{V}}}{e^2 } 
 \overline{\sigma} ( {\bf{q}} , \omega )
 \; .
 \label{eq:Pisigmaexact}
 \end{equation} 
From Eq.(\ref{eq:sigma2av}) we thus obtain
 \begin{equation}
 \overline{\sigma}^{(2)} ( {\bf{q}} , \omega , \phi ) =
 \frac{c}{2} \frac{ {\cal{V}}}{ (- i \omega )} \frac{ \partial}{\partial \phi}
 \overline{\sigma} ( {\bf{q}} , \omega )
 \; .
 \label{eq:sigma2avnew}
 \end{equation}
At finite ${\bf{q}}$, the dynamic conductivity has a diffusion
pole\cite{Pines89}. In fact, according to Ref.\cite{Vollhardt80}
in the limit of small wave-vectors and frequencies
 \begin{equation}
 \overline{\sigma} ( {\bf{q}} , \omega ) =
 \frac{ i \omega}{ i \omega - {\cal{D}} ( \omega ) {\bf{q}}^2 } 
 \overline{\sigma} ( \omega )
 \; , 
 \label{eq:sigmaapprox}
 \end{equation}  
where $\overline{\sigma} ( \omega) = \overline{\sigma} ( 0 , \omega)$,
and the frequency-dependent diffusion coefficient ${\cal{D}} (
\omega)$
is related to the dynamic conductivity via\cite{Vollhardt80}
 \begin{equation}
 \frac{ {\cal{D}} ( \omega)}{ {\cal{D}}_0} =
 \frac{ \overline{\sigma} ( \omega)}{\sigma_0}
 \; .
 \label{eq:Domega}
 \end{equation}
Here  ${\cal{D}}_0$ is the classical diffusion coefficient, which
is related to the Drude conductivity $\sigma_0$ via the
Einstein relation $ {\cal{D}}_0 = ( \Delta {\cal{V}} )^{-1} e^2
\sigma_0$.
We thus conclude
 \begin{equation}
 \overline{\sigma}^{(2)} ( {\bf{q}} , \omega , \phi ) =
 \frac{c}{2}  {\cal{V}} \frac{ \partial}{\partial \phi}
 \left[
\frac{  \overline{\sigma} (  \omega ) }{  {\cal{D}} ( \omega ) 
{\bf{q}}^2 - i \omega  } \right] 
 \; .
 \label{eq:sigma2avnew2}
 \end{equation}
Diagrammatically, the
diffusion pole in Eqs.(\ref{eq:sigmaapprox}) and (\ref{eq:sigma2avnew2})
implicitly takes the so-called {\it{diffuson}} diagrams into 
account\cite{Altshuler85}.
On the other hand, the weak-localization corrections
described by the {\it{Cooperon}} diagrams have to be 
included explicitly in the calculation $\overline{\sigma} ( \omega)$
and ${\cal{D}} ( \omega)$.  These diagrams are responsible for the dominant
dependence on the magnetic flux.

\subsection{Averaging over disorder}
\label{subsec:Averaging}

According to Eqs.(\ref{eq:Domega}) and (\ref{eq:sigma2avnew2})
the average non-linear conductivity can be expressed in terms of the
flux-dependent part of the average linear conductivity.
The latter
is determined by the famous weak localization correction
arising from coherent backscattering\cite{Altshuler85},
 \begin{equation}
 \frac{\partial}{\partial \phi}  \overline{\sigma} ( \omega)
 = 
 - \frac{e^2 {\cal{D}}_0 }{\pi {\cal{V}}} \frac{ \partial}{\partial \phi}
 {\sum_{  {\bf{Q}}  }}^{\prime}
 \frac{ 1}{ {\cal{D}}_0  ({\bf{Q}} + 2 {\bf{a}})^2 - i {\omega}  + {\cal{D}}_0 / L_{\varphi}^2 }
 \;  ,
 \label{eq:gWL}
 \end{equation}
where the prime means that
the sum is restricted to 
$| {\bf{Q}} | 
\raisebox{-0.5ex}{$\; \stackrel{<}{\sim} \;$} 2 \pi /\ell$,
and $L_{\varphi}$ is the dephasing length\cite{Altshuler85}.
Scaling out the Thouless energy $E_c = {\cal{D}}_0 / L^2$ and setting
now ${\bf{q}} = 0$ in Eq.(\ref{eq:sigma2avnew2}), 
we 
obtain from Eq.(\ref{eq:gWL})
 \begin{equation}
 \overline{\sigma}^{(2)} ( \omega , \phi ) 
 = \frac{c E_c ( e L / E_c)^2}{  ( - i
 \bar{\omega} ) \phi_0 }
 g ( \omega , \phi )
 \; , 
 \label{eq:sigma2scale}
 \end{equation}
where $\bar{\omega} = \omega / E_c$, and the dimensionless function
$g ( \omega , \phi )$ is given by
 \begin{equation}
 g ( \omega , \phi ) = - \frac{ \phi_0}{2 \pi}
 \frac{ \partial}{\partial \phi} 
 {\sum_{  {\bf{Q}}  }}^{\prime}
 \frac{ E_c}{ {\cal{D}}_0  ({\bf{Q}} + 2 {\bf{a}})^2 - i {\omega}  
 + {\cal{D}}_0 / L_{\varphi}^2 }
 \;  .
 \label{eq:gWL2}
 \end{equation}
For a thin ring with $L_{\bot} 
\raisebox{-0.5ex}{$\; \stackrel{<}{\sim} \;$} \ell$ the
${\bf{Q}}$-summation
in Eq.(\ref{eq:gWL2}) is one-dimensional, and can be carried out
exactly, with the result
 \begin{eqnarray} 
 g ( \omega , \phi )
 = \frac{2 e^{-W}}{W} 
 \frac{ 
 \sin ( 4 \pi \phi / \phi_0)
 [1 - e^{-  2 W}]}{ \left[ 1 - 2 e^{-W} \cos ( 4 \pi \phi /
 \phi_0) + e^{- 2 W} \right]^2} 
 \; .
 \label{eq:gsum}
 \end{eqnarray}
Here $W = \sqrt{ ( L / L_{\varphi})^2 - i \bar{\omega}}$, where the root has
to be taken such that ${\rm Re} W \geq 0$. For $| W | \ll 1$ this reduces to
 \begin{equation}
 g ( \omega , \phi) = \frac{ 4 \sin ( 4 \pi \phi / \phi_0)}{
\left[  4  \sin^2 ( 2 \pi \phi /
  \phi_0) - i \bar{\omega} + ( L / L_{\varphi})^2
 \right]^2}
 \; .
 \label{eq:gsimplify}
 \end{equation}
Note that by definition  $L_{\varphi} \gg L$ in a mesoscopic system, so
that for $| \omega | \ll E_c$ the parameter
$| W |$ is smaller than unity.
On the other hand, for
$| \omega | 
\raisebox{-0.5ex}{$\; \stackrel{>}{\sim} \;$} E_c$
the prefactor $e^{-W}$ in Eq.(\ref{eq:gsum})
reduces to $\exp [ -  \sqrt{| \bar{\omega} | /  2 }]$, so that
the non-linear conductivity becomes exponentially
small.
We  disagree in this point with Kravtsov and
Yudson\cite{Kravtsov93}, who found  
that the non-linear conductivity 
in the regime $E_c \ll \omega \ll \tau^{-1}$
is finite and approximately frequency-independent.
In view of the close connection between the non-linear conductivity
and the
Fock contribution to the equilibrium persistent current discussed
above, 
we think that our result is physically more reasonable.
The fact that for $| \omega | \gg E_c$ the non-linear conductivity
(\ref{eq:sigma2scale}) is exponentially small
is closely related to the exponential suppression of the contribution
from Matsubara frequencies larger than $E_c$ to the average Fock current
$\overline{I}_F$ in 
Eq.(\ref{eq:IF}).

\section{Possible experimental tests}
\label{sec:conc}

For simplicity, let us consider a time-dependent but spatially constant
external electric field along the circumference of the ring,
 \begin{equation}
 E ( t ) =  E ( \omega ) \cos ( \omega t )
 \; .
 \label{eq:field}
 \end{equation}
Because at zero wave-vector and finite frequencies the 
polarization  $\overline{\Pi} ( 0 , \omega
)$ vanishes, this
field is not screened. Hence, 
$ {\hat{\bf{q}}} \cdot {\bf{E}} ( {\bf{q}} = 0, \omega)$
in Eq.(\ref{eq:Ineav}) can be identified with the
external field $E ( \omega)$.
Experimentally, the field (\ref{eq:field}) can be generated
by a time-dependent magnetic flux 
through the center of the ring,
 \begin{equation}
 \phi ( t )  = \phi + \phi ( \omega ) \sin ( \omega t )
 \; .
 \label{eq:phiomega}
 \end{equation}
By Faraday's law of induction, the relation between $E ( \omega)$
and $\phi ( \omega)$ is
 \begin{equation}
 e L E ( \omega ) = 2 \pi \omega \frac{\phi ( \omega ) }{ \phi_0 }
 \;  .
 \label{eq:Faraday}
 \end{equation}
Note that in the experiment\cite{Levy90} the current
was measured in the presence of such a 
time-dependent flux with
frequencies in the range
between $10$ and $10^3 {\rm Hz}$. In this  range
no  frequency-dependence of the current was detected, so that
apparently the measurements were performed in the static limit.
Note, however, that in principle one should distinguish
between the thermodynamic equilibrium current that is
determined by the flux-dependent part of the free energy, and
the dynamic current that is obtained from the 
time-dependent response in the limit of vanishing
frequency\cite{Efetov91}.
In the present work we are interested in the
frequency range $\Delta \ll \omega \ll \tau^{-1}$, corresponding
to frequencies between $10^{8}$ and $10^{13} {\rm Hz}$. 
We  predict that the static
non-equilibrium current should become exponentially small
as soon as the frequency of the electric field
exceeds the Thouless energy $E_c $.
{\it{This effect can be used to directly measure the Thouless energy of a
mesoscopic ring.}}
For the rings used in Ref.\cite{Levy90} the time-dependent
non-equilibrium current should disappear for $\omega \approx 10^{10}
{\rm Hz}$. We would like to emphasize that this prediction
can be verified without any modifications of the experimental
setup used in Refs.\cite{Levy90,Mohanty96}.

\begin{figure}
\epsfysize6cm 
\hspace{5mm}
\epsfbox{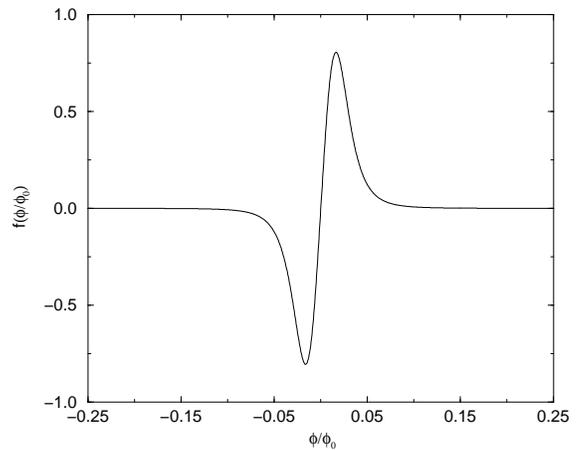}
\vspace{5mm}
\caption{
$f ( \bar{\phi} ,\bar{\omega} , \gamma)$
as function of ${\phi} / \phi_0$ for $\bar{\omega} = 0.1 $ and $\gamma = 1$.
}
\label{fig:graph}
\end{figure}

\begin{figure}
\epsfysize6cm 
\hspace{5mm}
\epsfbox{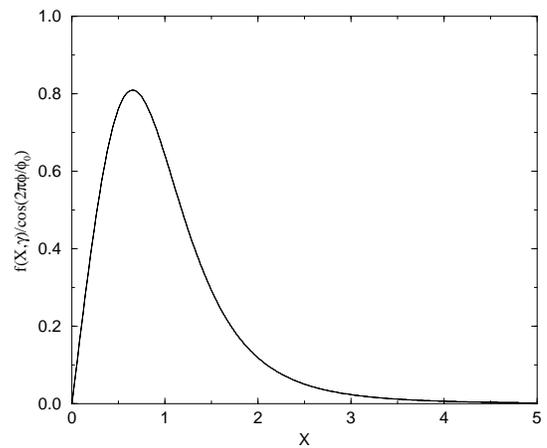}
\vspace{5mm}
\caption{$\tilde{f} ( X , \gamma ) = 
f / \cos ( 2 \pi \bar{\phi})$ as function of
$X$ (see Eq.(\ref{eq:Xdef})) for $\gamma = 1$.
}
\label{fig:fX}
\end{figure}

\begin{figure}
\epsfysize6cm 
\hspace{5mm}
\epsfbox{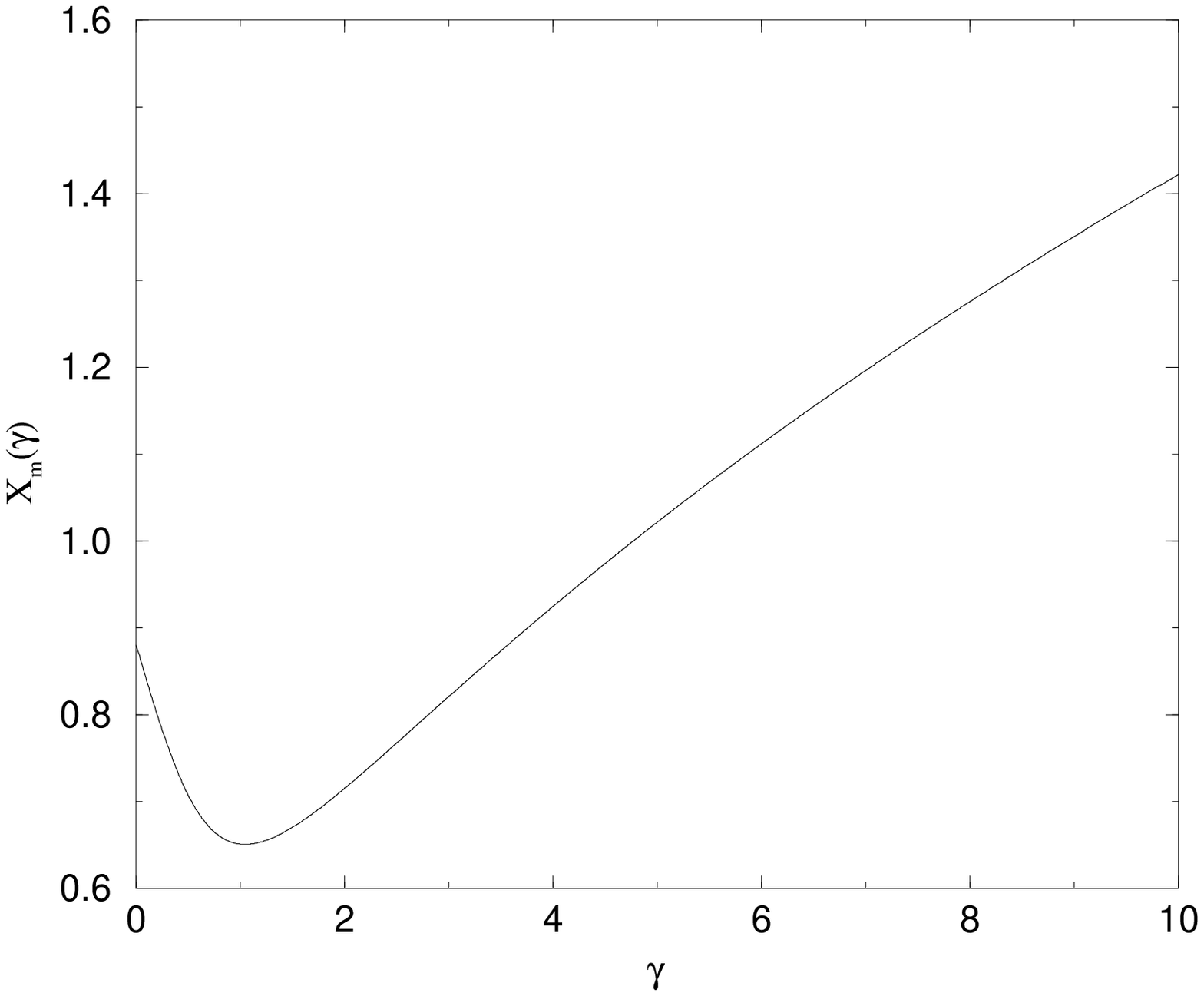}
\vspace{5mm}
\caption{
Position of the maximum $X_m ( \gamma )$ of $\tilde{f} ( X , \gamma)$ as function
of $\gamma$. For $\gamma = 0$ it is easy to show that
$ X_m ( 0) = ( 3/5)^{1/4} \approx 0.88$, while 
$X_m ( \gamma ) \propto \sqrt{\gamma}$ for $\gamma
\rightarrow \infty$. 
}
\label{fig:fmax}
\end{figure}

Because the non-equilibrium current 
$\overline{I}^{(2)}$ is driven by an external time-dependent
flux $\phi ( \omega)$, it can be easily distinguished
from the thermodynamic equilibrium current.
Let us now discuss the expected size of this non-equilibrium current. 
Given the fact that the external field in Eq.(\ref{eq:field})
has a
$\cos$-dependence, the
experimentally measured non-equilibrium current 
is determined by the real part of the non-linear conductivity,
 \begin{equation}
 \overline{I}^{(2)} = {\rm Re} 
 \left[ \overline{\sigma}^{(2)} ( \omega , \phi ) \right] 
 \left[   \frac{  \omega }{ e L } \right]^2
 \left[   \frac{ 2 \pi \phi ( \omega)}{\phi_0} \right]^2 
 \; .
 \label{eq:I2omegafinal}
 \end{equation}
For frequencies $| \omega |
\raisebox{-0.5ex}{$\; \stackrel{<}{\sim} \;$} E_c$
we may use Eq.(\ref{eq:gsimplify}) to simplify the
non-linear conductivity, so that in this regime we obtain
after some rescalings
 \begin{equation}
  \overline{I}^{(2)}  \approx
 - \frac{ c E_c}{\phi_0 } 
 \left[   \frac{ 2 \pi \phi ( \omega)}{\phi_0}
 \right]^2  | \bar{\omega} |^{-1/2} f ( \bar{\phi } , \bar{\omega } ,
 \gamma )
 \label{eq:Ifinal}
 \; ,
 \end{equation}
where $\bar{\phi} = \phi / \phi_0$,
$\bar{\omega} = \omega / E_c$, $\gamma = {\cal{D}}_0 / ( L_{\varphi}^2
| \omega | )$.
Defining the dimensionless variable
 \begin{equation}
 X = 2 \frac{ \sin ( 2 \pi \bar{\phi})}{\sqrt{ | \bar{\omega}| }}
 \; ,
 \label{eq:Xdef}
 \end{equation}
the function $f ( \bar{\phi} , \bar{\omega} ,
\gamma)$ can be written as
 \begin{equation}
 f \left( \bar{\phi} ,\bar{\omega} , \gamma \right)
 =  8   \cos ( 2 \pi \bar{\phi}) 
 \frac{ X [ \gamma + X^2 ]}{ \left[ 1 + \left[ \gamma + X^2 \right]^2
 \right]^2}
 \label{eq:fX}
 \; .
 \end{equation}
A graph of $ f ( \bar{\phi} , \bar{\omega} , \gamma)$
as function of $\bar{\phi} = \phi / \phi_0$
for $\bar{\omega} = 0.1$ and $\gamma = 1$ is shown in
Fig.\ref{fig:graph}.
%
%
%
%
Obviously the size of the current
$\overline{I}^{(2)}$ in Eq.(\ref{eq:Ifinal}) is determined by
three experimentally controllable parameters: $\phi ( \omega)$ ,
$\phi$, and $\omega$.
Let us find the values of these parameters that 
maximize the current.
Obviously $\phi ( \omega)$ should be chosen as large as possible. 
It should be kept in mind, however, that Eq.(\ref{eq:Ifinal}) 
is the quadratic order in a systematic expansion in powers of
$\phi ( \omega) / \phi_0$.
Higher orders should be negligible as long as
$| \phi (\omega)  | \ll  \phi_0$.
Thus,
the largest value of $| \phi ( \omega) |$ where Eq.(\ref{eq:Ifinal}) can be expected to be
accurate is  
 \begin{equation}
 | \phi ( \omega ) | \approx \frac{ \phi_0 }{  2 \pi}
 \; .
 \label{eq:phiomegaopt}
 \end{equation}
Next, consider the optimal choice of the frequency.
Because of the factor of $|\bar{\omega}|^{-1/2}$ in
Eq.(\ref{eq:Ifinal}),
it is advantageous to choose the frequency as small as possible.
However, our perturbative
calculation breaks down for frequencies of the order of the
mean level spacing $\Delta$. The optimal choice is therefore
 \begin{equation}
 \omega \approx \Delta 
 \; .
 \label{eq:omegaopt}
 \end{equation}
Finally, from Eq.(\ref{eq:fX}) and Figs.\ref{fig:fX},
\ref{fig:fmax} it is clear that the
function $ \tilde{f} ( X , \gamma ) \equiv f / \cos ( 2 \pi \bar{\phi})$ has a maximum
at a value $X_m ( \gamma)  = O (1)$ (provided $\gamma $ is not too large). 
This implies that for $| \omega | \ll E_c$ the non-equilibrium current
is maximal at the static flux $\phi = \phi_m$, where
 \begin{equation}
 \phi_m = \frac{\phi_0}{4 \pi} \sqrt{ \frac{|\omega |}{E_c}}
 X_m \left( \frac{ {\cal{D}}}{ L_{\varphi}^2 | \omega | } \right)
 \; .
 \label{eq:phistatopt}
 \end{equation}
The function $X_m ( \gamma)$ is shown in Fig.\ref{fig:fmax}.
%
%
%
%
%
%
%
%
From Eq.(\ref{eq:Xdef})  we also see that 
the width $\Delta \phi$ of the interval
around $\phi_m$ (modulo  $\phi_0 /2$)
where the current is enhanced is 
$\Delta \phi \approx \phi_0 | \bar{\omega}|^{1/2} / (4 \pi)$. 
Outside this interval,
which is rather narrow for $| \bar{\omega} | \ll 1$,
the non-equilibrium current is much smaller than 
at the maxima (see Fig.\ref{fig:graph}).  
In fact, for $ | \phi - \phi_m | \gg \Delta \phi$
the parameter $| X |$ in Eq.(\ref{eq:Xdef}) 
is large compared with unity, so that 
we may approximate
 \begin{equation}
 f \left( \bar{\phi} ,\bar{\omega} , \gamma \right)
 \approx     \frac{| \bar{\omega} |^{5/2}}{4} \frac{ \cos ( 2 \pi \bar{\phi}) }{
 \sin^5 ( 2 \pi \bar{\phi}) }
 \; \; \; ,
 \; \;  | X | \gg 1
 \; ,
 \label{eq:fXapprox}
 \end{equation}
where we have assumed that $\gamma \raisebox{-0.5ex}{$\; \stackrel{<}{\sim} \;$} 1$.
Using $2 \sin^2 (x) =  [ 1 - \cos(2x)]$, the
non-equilibrium current in this regime can be written
as
  \begin{equation}
  \overline{I}^{(2)} \approx
 - \frac{ c E_c}{\phi_0 } 
 \left[   \frac{ 2 \pi \phi ( \omega)}{\phi_0}
 \right]^2  | \bar{\omega} |^2
 \frac{  \sin ( 4 \pi \bar{\phi})}{ [1 - \cos ( 4 \pi \bar{\phi})]^3}
 \label{eq:Ifinal2}
 \; .
 \end{equation}
For the parameters of the experiment\cite{Levy90}
(taking now the spin degeneracy into account), we find that at the
optimal values of the parameters given
in Eqs.(\ref{eq:phiomegaopt}--\ref{eq:phistatopt})
the maximal amplitude of the 
non-equilibrium
current
is $\overline{I}^{(2)}_{\rm max} \approx 5 \times 10^{-3} e v_F / L$.
This current has the same order of magnitude as the equilibrium
current
measured in Refs.\cite{Levy90,Mohanty96},
and therefore should be measurable with the available technology.
The rather pronounced peaks of $\overline{I}^{(2)}$
as function of $\phi$
for small values of $\phi$ (modulo $\phi_0 /2$)
distinguish the non-equilibrium current
from the thermodynamic persistent current.

\section{Summary}
\label{sec:sum}

In this paper we have presented a thorough theoretical analysis of the
static non-equilibrium current $I^{(2)}$ 
in a mesoscopic metal ring threaded by a magnetic flux,
which is generated in quadratic response to a
time-dependent electric field.  
Using a path integral approach, we have
shown that this non-equilibrium current is closely related to the
screened exchange contribution to the thermodynamic persistent
current\cite{Ambegaokar90,Kopietz97b}.
In fact, from Eqs.(\ref{eq:IF}), (\ref{eq:IFrewrite}), and
(\ref{eq:sigmanldef})
it is obvious that the 
weight of the RPA screened interaction
in the exchange correction to the equilibrium persistent
current
at fixed energy-momentum transfer is
essentially given by the non-linear conductivity
$\sigma^{(2)} ( \omega , \phi )$ 
associated with $I^{(2)}$.
This observation has allowed us to derive the relation (\ref{eq:sigma2av})
between the non-linear conductivity and the
flux-dependent part of the polarization.
Thus, the disorder average $\overline{\sigma}^{(2)} ( \omega , \phi)$
of the non-linear conductivity is directly
related to the weak-localization 
correction to the average polarization, which in turn
can be expressed in terms of the weak-localization correction
to the linear conductivity. In other words,
{\it{the average non-linear conductivity is directly
related to the weak localization correction to the
linear conductivity}}.

A measurement of the average static non-equilibrium current
$\overline{I}^{(2)}$ in a mesoscopic
metal ring as function of frequency
and static flux would be very interesting for several reasons.
Such a measurement would directly probe the 
weak localization corrections to the frequency-dependent
polarization and the linear conductivity  of the system.
In contrast to Kravtsov and Yudson\cite{Kravtsov93}, we predict that
the static non-equilibrium current should be exponentially
suppressed when the external frequency exceeds the Thouless energy.
This is in agreement with the close connection
between the non-equilibrium current $I^{(2)}$ and 
the Fock contribution $I_F$ to the equilibrium persistent
current. The latter is known to be exponentially suppressed if the
temperature  becomes larger than the Thouless 
energy\cite{Ambegaokar90}. Note that in this case the temperature
acts as an infrared cutoff, just like the external
frequency in the case of the non-equilibrium current. 
Our theory can be verified experimentally from the
measurement of the {\it{time-independent}} non-linear current response
of a mesoscopic metal ring pierced by a {\it{time-dependent}}
external flux $\phi ( t )$ of the form (\ref{eq:phiomega}), with
frequencies in the range 
$ \Delta 
\raisebox{-0.5ex}{$\; \stackrel{<}{\sim} \;$}
\omega  \ll \tau^{-1}$. We hope that such an experiment
will be done in the near future.

\section*{Acknowledgement}

We would like to thank Kurt Sch\"{o}nhammer for his
constructive comments. 
This work was supported by the
Deutsche Forschungsgemeinschaft (SFB 345).

%

%
%

%


\begin{thebibliography}{99}
\bibitem{footnote2}
In a finite system the Fourier transform $f_{\bf{q}}$ of the Coulomb potential
does not have any singularities. 
It is then convenient to normalize $f_{\bf{q}}$ 
in Eq.(\ref{eq:Seffdef}) such that
$f_{\bf{q}} \sim e^{2} /C_0$ for ${\bf{q}} \rightarrow 0$, where
$C_0$ is the classical capacitance\cite{Kopietz93}.
Similarly, the polarization is normalized such that to leading order
$\overline{\Pi}_0 ( {\bf{q}} , 0 ) \sim \Delta^{-1}$.
\end{thebibliography}
\end{document}